\documentclass[twocolumn,showpacs,preprintnumbers]{revtex4}
\usepackage{latexsym}
\usepackage{graphicx}
\usepackage{dcolumn}
\usepackage{bm}
\usepackage{xcolor}

\input amssym.def
\input amssym.tex

\newcommand{\be}{\begin{equation}}
\newcommand{\ee}{\end{equation}}
\newcommand{\bear}{\begin{eqnarray}}
\newcommand{\eear}{\end{eqnarray}}
\newcommand{\ba}{\begin{array}}
\newcommand{\ea}{\end{array}}

\begin{document}

\title{Cardy-Verlinde formula in Taub-NUT/Bolt-(A)dS space}

\author{Chong Oh Lee}
 \altaffiliation{Department of Physics, Chonbuk National University, Jeonju
561-756, Republic of Korea}
\email{cohlee@chonbuk.ac.kr}

\date{\today}

\begin{abstract}
We consider a finite action for a higher dimensional
Taub-NUT/Bolt-(A)dS space via the so-called counter term subtraction
method. In the limit of high temperature, we show that the
Cardy-Verlinde formula holds for the Taub-Bolt-AdS metric and for
the specific dimensional Taub-NUT-(A)dS metric, except for the
Taub-Bolt-dS metric.
\end{abstract}
\maketitle

\setcounter{equation}{0}
\section{Introduction}
The AdS/CFT duality was first conjectured
by~\cite{'t Hooft:1973jz} in his search for relationship between
gauge theories and strings.
The AdS/CFT correspondence
\cite{Maldacena:1997re,Gubser:1998bc,Witten:1998qj,Aharony:1999ti,D'Hoker:2002aw}
asserts there is an
equivalence between a gravitational theory in the bulk and a conformal field theory in the
boundary.
According to AdS/CFT, a $(d+1)$-dimensional S-(A)dS action $A$ is given by
\bear\label{totadsact}
A=A_B+A_{\partial B}+A_{ct}
\eear where the bulk action $A_B$, action
boundary $A_{\partial B}$, and counterterm action $A_{ct}$ are given as
\bear\label{adsact}
A_B&=&\frac{1}{16\pi G_{d+1}}\int_{\cal M}d^{d+1}x\sqrt{-g}({\cal R}-2\Lambda),\nonumber\\
A_{\partial B}&=&-\frac{1}{8\pi G_{d+1}}\int_{\partial {\cal M}}d^d x
\sqrt{-\gamma}\Theta,\nonumber\\
A_{ct}&=&-\frac{1}{8\pi G_{d+1}}\int_{\partial {\cal M}} d^d x \sqrt{-\gamma}
\left\{-\frac{d-1}{l}\right.\nonumber\\ &&- \frac{l R}{2(d-2)}{\cal F}(d-3)\nonumber\\
&&-\frac{l^3}{2(d-2)^2(d-4)}\nonumber\\
&&\hspace{1cm}\times \left(R_{ab}R^{ab}-\frac{d}{4(d-1)}R^2\right)
{\cal F}(d-5)\nonumber\\
&&+\frac{l^5}{(d-2)^3(d-4)(d-6)}
\nonumber\\
&&\hspace{1cm}\times\left(\frac{3d+2}{4(d-1)}RR_{ab}R^{ab}
-\frac{d(d+2)}{16(d-1)^2}R^3\right.\nonumber\\
&&\left.\left.
+\frac{d-2}{2(d-1)}R^{ab}\nabla_a \nabla_b R -R^{ab}\Box R_{ab}\right.\right.
\nonumber\\
 &&\left.\left.+\frac{1}{2(d-1)}R\Box R
\right){\cal F}(d-7)+\cdots\right\},
\eear
where a negative cosmological constant $\Lambda$ is $\Lambda=-{d(d-1)}/{2l^2}$, $\Theta$
is the trace of extrinsic curvature.
Here, ${\cal F}(d)$ is a step function, 1 when $d\geq 0$, 0 otherwise.
The boundary action $A_{\partial B}$ is added to the action $A$ to obtain
equations of motion well behaved at the boundary.
Then the boundary energy-momentum tensor is expressed in \cite{Brown:1992br}
\bear\label{bdryenergymoment}
\frac{2}{\sqrt{-\gamma}}\frac{\delta A_{\partial B}}{\delta \gamma^{ab}}=\Theta_{ab}
-\gamma_{ab}\Theta.
\eear
The counterterm action $A_{ct}$ is added to the action $A$ to remove the divergence appearing
as the boundary goes to infinity \cite{Balasubramanian:1999re}.
For low dimensional S-AdS, a few terms in the
counterterm action $A_{ct}$ were explicitly evaluated in
\cite{Balasubramanian:1999re,Emparan:1999pm}. Using the universality of the structure of
divergences, the counterterm action $A_{ct}$ for arbitrary dimension is suggested
in \cite{Kraus:1999di}. This action $A$ (\ref{totadsact}) leads to the entropy $S$
via the Gibbs-Duhem relation
\bear\label{bentro}
S=\frac{E}{T}-A
\eear
where $T$ denotes the temperature and $E$ is the total energy.

The entropy of the (1+1)-dimensional CFT is expressed in terms of
the Virasoro operator $L_0$ and the central charge $c$,
the so-called the Cardy formula \cite{Cardy:1986ie}.
Using conformal invariance, the generalized Cardy formula in arbitrary dimension is shown
to be given universal form as~\cite{Verlinde:2000wg}
(for the review articles of the issue, see, e.g.,
\cite{Nojiri:2001fa,Nojiri:2002hz,Lidsey:2002ah})
\begin{equation}\label{cv}
S_{\rm CFT}=\frac{2\pi R}{\sqrt{ab}}\sqrt{E_c(2E-E_c)}\,,
\end{equation}
where $a$ and $b$ are certain constants.
$R$ denotes the radius of the universe at a given time
and $E_c$ is the Casimir energy defined by
\begin{equation}\label{Ec}
E_c=d\,E-(d-1)TS\,.
\end{equation}
Employing AdS/CFT dual picture,
$\sqrt{ab}$ is fixed to $(d-1)$ exactly, in particular,
for a $d$-dimensional CFT on ${\rm \mathbf{R}}$$\times S^{d-1}$
~\cite{Verlinde:2000wg}.
Then, the entropy is given as
\begin{equation}\label{cv0}
S_{\rm CFT}=\frac{2\pi R}{d-1}\sqrt{E_c(2E-E_c)}\,,
\end{equation}
which is shown to hold for Schwarzschild (A)dS
(S-(A)dS)~\cite{Verlinde:2000wg,Cai:2001sn}, charged
(A)dS~\cite{Cai:2001jc,Cai:2001tv},
Kerr-(A)dS~\cite{Klemm:2001db,Cai:2001tv}, and
Taub-Bolt-AdS$_4$~\cite{Birmingham:2001vd}. There are many other
relevant papers on the subject
\cite{Wang:2001bf,Wang:2001bv,Setare:2002ss,Setare:2002qa,Setare:2003fg}.
Thus, one may naively expect that the entropy of all CFTs that have
an AdS-dual description is given as the form (\ref{cv0}). However,
AdS black holes do not always satisfy the Cardy-Verlinde formula
(see, e.g.,~\cite{Cai:2001jc,Gibbons:2005vp}). Therefore, one
intriguing question is whether this formula is valid for higher
dimensional Taub-NUT-(A)dS at high temperature. In this Letter, we will
endeavor to do this.

\section{Taub-NUT/Bolt-AdS  black hole}
When the total number of dimension of the spacetime is even,
$(d+1)=2u+2$, for some integer $u$, the Euclidean section of the
arbitrary (d+1)-dimensional-Taub-NUT-AdS metric, for a $U(1)$
fibration over a series of the space ${\cal M}^2$ as the base space
$\bigotimes_{i=1}^{u}{\cal M}^2$, is given by
\cite{Chamblin:1998pz,Hawking:1998ct,Mann:1999bt,Awad:2000gg,
Clarkson:2002uj,Astefanesei:2004ji,Astefanesei:2004kn}
(for the generalized versions of the issue, see, e.g.,
\cite{Mann:2003zh,Mann:2005ra})
\bear
ds^2_{\rm AdS}&=&f(r)\left[dt_{E}
+2N\sum_{i=1}^{u}\cos(\theta_i)d\phi_i\right]^2
\nonumber\\&+&\frac{dr^2}{f(r)}
+(r^2-N^2)\sum_{i=1}^{u}
\Biggr[d\theta_i^2+\sin^2(\theta_i)d\phi_i^2\Biggr],
\nonumber\\
\eear where $N$ represents a NUT charge for the Euclidean section,
and the metric function $f(r)$ has the general form
\bear
f(r)&=&\frac{r}{(r^2-N^2)^u}\int^{r}
\left[\frac{(p^2-N^2)^u}{p^2}\right.
\nonumber\\&+&\left.\frac{(2u+1)(p^2-N^2)^{u+1}}{l^2 p^2}
\right]dp-\frac{2mr}{(r^2-N^2)^u},
\eear
with a cosmological parameter $l$ and a geometric mass $m$.

Requiring $f(r)|_{r=N}$ the NUT solution occurs. Then
for AdS spacetime the inverse of the temperature $\beta$
arises from imposed condition in order to ensure regularity in
the Euclidean time $t_{E}$ and radial coordinate $r$
\cite{Chamblin:1998pz,Hawking:1998ct,Mann:1999bt,Awad:2000gg,
Clarkson:2002uj,Astefanesei:2004ji,Astefanesei:2004kn}
\bear\label{HT0}
\beta=\left.\frac{4\pi}{f'(r)}\right|_{r=N}=
\frac{2(d+1)\pi N}{q},
\eear
where $\beta$  is the period of $t_{E}$. Here
$q$ is a positive integer, which originates from removing Misner string singularities.
Using counter term subtraction method the regularized action is given as
\cite{Chamblin:1998pz,Hawking:1998ct,Mann:1999bt,Awad:2000gg,
Clarkson:2002uj,Astefanesei:2004ji,Astefanesei:2004kn}
\bear\label{ac0}
I_{\rm NUT}&=&\frac{(4\pi)^{\frac{d}{2}}N^{d-2}
\biggr((d-1)N^2-l^2\biggr)}
{32\pi^{2}l^2}\nonumber\\
&&\hspace{3mm}\times\Gamma(\frac{2-d}{2})\Gamma(\frac{d+1}{2})\beta.
\eear
Employing the thermal relation $E= \partial_{\beta} I$
the total energy can also be written by
\bear\label{e0}
E&=&\frac{(4\pi)^{\frac{d}{2}}(d-1)N^{d-2}\biggr((d+1)N^2-l^2\biggr)}
{32\pi^2ql^2}\nonumber\\
&&\hspace{3mm}\times\Gamma(\frac{2-d}{2})\Gamma(\frac{d+1}{2}),
\eear
and the entropy is given as
\cite{Chamblin:1998pz,Hawking:1998ct,Mann:1999bt,Awad:2000gg,
Clarkson:2002uj,Astefanesei:2004ji,Astefanesei:2004kn}
\bear\label{s0}
S_{\rm NUT, AdS}&=&\frac{(4\pi)^{\frac{d}{2}}N^{d-2}\biggr(d(d-1)N^2-(d-2)l^2\biggr)
}{32\pi^{2}l^2}
\nonumber\\
&&\hspace{3mm}\times\Gamma(\frac{2-d}{2})\Gamma(\frac{d+1}{2})\beta,
\eear
by the Gibbs-Duhem relation $S=\beta M - I$
where $M$ denotes the conserved mass
\bear
M=\frac{(d-1)(4\pi)^{\frac{d}{2}}}{16\pi^{\frac{3}{2}}}m.
\eear
Substituting (\ref{HT0}), (\ref{e0}), and (\ref{s0}) into (\ref{Ec}), one gets
the Casimir energy~\cite{Verlinde:2000wg}
\bear
E_{c}&=&\frac{(4\pi)^{\frac{d}{2}}(d-1)N^{d-2}\biggr(dN^2-l^2\biggr)}
{16\pi^2ql^2}\nonumber\\
&&\hspace{3mm}\times\Gamma(\frac{2-d}{2})\Gamma(\frac{d+1}{2}).
\eear
From now on, for convenience
we use $l/z$ instead of the universe radius $R$ in (\ref{cv0})
since the AdS metric is always asymptotically taken to be~\cite{Fefferman:1984aa}
\bear\label{FGseries}
ds^2=\frac{l^2}{z^2}dz^2+\frac{l^2}{z^2}g_{ab}(x, z)dx^a dx^b,
\eear
where the $r=\infty$ is put to $z=0$, and
the roman indexes $a$ and $b$ refer to boundary coordinates.
When $1/\sqrt{ab}$ in the formula (\ref{cv0}) is taken to
be $2/(d+1)(d-1)(d-2)$, the CFT entropy is given as
\bear
S_{\rm CFT}&=&\frac{4\pi l \sqrt{|E_c(2E-E_c)|}}{(d+1)(d-1)(d-2)}\,,
\nonumber\\
&=&\frac{(4\pi)^{\frac{d}{2}}|dN^2-l^2|(-1)^{[\frac{d}{2}]}
\Gamma(\frac{d+1}{2})\Gamma(\frac{2-d}{2})}{4\pi (d+1)(d-2)q},
\eear
where $[x]$ is the Gauss number (greatest integer less than or
equal to x).
Here it seems that
the difference from the standard Cardy-Verlinde formula (\ref{cv0})
is due to the distinctive nature of NUT solution in AdS space
like asymptotically locally AdS (ALAdS) metric.
In the limit of high temperature, $N\rightarrow0$,
leading term in the entropy of CFT can be expressed as
\bear
S_{\rm CFT}&=&\frac{(4\pi)^{\frac{d}{2}}(d+1)(d-2)N^{d-1}}
{16\pi q}\nonumber\\
&&\hspace{3mm}\times (-1)^{[\frac{d}{2}]}
\Gamma(\frac{d+1}{2})\Gamma(\frac{2-d}{2})\nonumber\\
&=&(-1)^{[\frac{d}{2}]}S_{\rm NUT, AdS}.
\eear
This result shows that the entropy of
the Taub-NUT-AdS space suffices to be the generalized Cardy-Verlinde formula
(\ref{cv}) for all even $u$ $(d+1=2u+2)$.
This is reasonable because
the Taub-NUT-AdS metric has the thermodynamically stable range depending
on the magnitude of the NUT charge i.e. any NUT solution in AdS space for all odd $u$ is
thermodynamically unstable in the lime $N\rightarrow0$.

Requiring $f(r)|_{r=r_{\rm B}>N}$ and $f'(r)|_{r=r_{\rm B}}=\frac{1}{(u+1)N}$,
the Bolt solution occurs. In Taub-Bolt-AdS metric, the inverse of the temperature,
the total energy, and the entropy are respectively
\bear\label{HT1}
\beta=\left.\frac{4\pi}{f'(r)}\right|_{r=r_{\rm B}}=
\frac{4\pi l^2 r_{\rm B}}{l^2+(2u+1)(r_{\rm B}^2-N^2)},
\eear
\bear\label{e1}
E&=&\frac{(4\pi)^u\,u}{8\pi}\left(\sum_{k=0}^{u}\left(
\begin{array}{l}
u\\
k
\end{array}
\right)
\frac{(-1)^k N^{2k}r_{\rm B}^{2u-2k-1}}{2u-2k-1}
\right.
\nonumber\\
&&\hspace{5mm}+\left.\sum_{k=0}^{u+1}\left(
\begin{array}{l}
u+1\\
k
\end{array}
\right)\frac{(-1)^k N^{2k}r_{\rm B}^{2u-2k-1}}{2u-2k-1}\right),
\eear
\bear\label{s1}
S_{\rm Bolt, AdS}&=&\frac{(4\pi)^u\beta}{16\pi l^2}
\left[\frac{(2u-1)(2u+1)(-1)^uN^{2u+2}}{r_{\rm B}}\right.
\nonumber\\
&&\hspace{10mm}+\sum_{k=0}^{u}\left(
\begin{array}{l}
u\\
k
\end{array}
\right)(-1)^k N^{2k}r_{\rm B}^{2u-2k}
\nonumber\\&&\hspace{17mm}\times\left(\frac{(2u-1)l^2}{(2u-2k-1)r_{\rm B}}\right.
\nonumber\\&&+\left.\left.\frac{(2u+1)(2u^2+3u-2k+1)r_{\rm B}}{(2u-2k+1)(u-k+1)}
\right)\right],\nonumber\\
\eear
where $r_{\rm B}=\frac{ql^2+\sqrt{q^2l^4+(2u+1)(2u+2)^2N^2[(2u+1)N^2-l^2]}}
{(2u+1)(2u+2)N}$. The CFT entropy is written as
\bear
S_{\rm CFT}=\frac{2\pi l\sqrt{E_{\rm c}(2E-E_{\rm c})}}{2u\sqrt{2u-1}},
\eear
where $1/\sqrt{ab}$ is fixed to $1/2u\sqrt{2u-1}$.
In the high temperature limit, the CFT entropy well
suffices to be Cardy-Verlinde formula as the following
\bear
S_{\rm CFT}=\frac{(4\pi)^u}{4N^{2u}}
\left(\frac{ql^2}{2u^2+3u+1}\right)^{2u}=S_{\rm Bolt,AdS}.
\eear
Note that the higher dimensional Taub-Bolt-AdS space follows the
generalized Cardy-Verlinde formula (\ref{cv})
even if Taub-Bolt-AdS$_4$ space $(u=1)$ exactly satisfies the Cardy-Verlinde
formula (\ref{cv0})~\cite{Birmingham:2001vd}.

\section{Taub-NUT/Bolt-dS black hole}
Taub-NUT-dS metric is obtained from the Taub-NUT-AdS metric by
replacing $l^2\rightarrow-l^2$, and one has
\cite{Clarkson:2003wa,Clarkson:2003kt,Mann:2004mi}
\bear ds^2_{\rm
dS}&=&-g(r)\left[dt_{E}
+2N\sum_{i=1}^{u}\cos(\theta_i)d\phi_i\right]^2
\nonumber\\&-&\frac{dr^2}{g(r)} +(r^2-N^2)\sum_{i=1}^{u}
\Biggr[d\theta_i^2+\sin^2(\theta_i)d\phi_i^2\Biggr],
\nonumber\\
\eear
where $g(r)$ is given as
\bear
g(r)&=&-\frac{r}{(r^2-N^2)^u}\int^{r}
\left[\frac{(p^2-N^2)^u}{p^2}\right.
\nonumber\\&-&\left.\frac{(2u+1)(p^2-N^2)^{u+1}}{l^2 p^2}
\right]dp+\frac{2mr}{(r^2-N^2)^u}.
\eear
Using parallel way as in the previous case,
the inverse of the temperature, the total energy,  the entropy, and
the Casimir energy are obtained
\bear
\beta=\left.\frac{4\pi}{g'(r)}\right|_{r=N}=\frac{2(d+1)\pi |N|}{q},
\eear
\bear
E&=&\frac{(4\pi)^{\frac{d}{2}}(d-1)N^{d-2}\biggr((d+1)N^2+l^2\biggr)
}{32\pi^2ql^2}\nonumber\\
&&\hspace{3mm}\times\Gamma(\frac{2-d}{2})\Gamma(\frac{d+1}{2}),
\eear
\bear\label{NdSe}
S_{\rm NUT, AdS}&=&\frac{(4\pi)^{\frac{d}{2}}N^{d-2}\biggr(d(d-1)N^2+(d-2)l^2\biggr)
}{32\pi^{2}l^2}
\nonumber\\
&&\hspace{3mm}\times\Gamma(\frac{2-d}{2})\Gamma(\frac{d-1}{2})\beta,
\eear
\bear
E_{c}&=&\frac{(4\pi)^{\frac{d}{2}}(d-1)N^{d-2}\biggr(dN^2+l^2\biggr)}
{16\pi^2ql^2}\nonumber\\
&&\hspace{3mm}\times\Gamma(\frac{2-d}{2})\Gamma(\frac{d+1}{2}).
\eear
In this case, $1/\sqrt{ab}$ in (\ref{cv0}) is fixed to $2/(d+1)(d-1)(d-2)$.
Then, the entropy in boundary CFT is expressed as
\bear
S_{\rm CFT}&=&\frac{4\pi l\sqrt{E_c(2E-E_c)}}{(d+1)(d-1)(d-2)}\,,
\nonumber\\
&=&\frac{(4\pi)^{\frac{d}{2}}(dN^2+l^2)(-1)^{[\frac{d}{2}]}
\Gamma(\frac{d+1}{2})\Gamma(\frac{2-d}{2})}{4\pi (d+1)(d-2)q}.
\eear
In high temperature limit, leading term in the entropy of the CFT for all even $u$
is precisely matched with that in the entropy of the Taub-NUT-dS space
as the following
\bear
S_{\rm CFT}&=& \frac{(4\pi)^{\frac{d}{2}}(d+1)(d-2)N^{d-1}
\Gamma(\frac{4-d}{2})\Gamma(\frac{3+d}{2})}
{16\pi^{\frac{3}{2}}q}
\nonumber\\
&=& S_{\rm NUT, dS}.
\eear

For the Bolt solution in dS space, the inverse of the temperature,
the total energy, and the entropy are respectively
\bear\label{HT1}
\beta=\left.\frac{4\pi}{f'(r)}\right|_{r=r_{\rm B}}=-
\frac{4\pi l^2 r_{\rm B}}{l^2+(2u+1)(r_{\rm B}^2-N^2)},
\eear
\bear\label{e1}
E&=&-\frac{(4\pi)^u\,u}{8\pi}\left(\sum_{k=0}^{u}\left(
\begin{array}{l}
u\\
k
\end{array}
\right)
\frac{(-1)^k N^{2k}r_{\rm B}^{2u-2k-1}}{2u-2k-1}
\right.
\nonumber\\
&&\hspace{5mm}+\left.\sum_{k=0}^{u+1}\left(
\begin{array}{l}
u+1\\
k
\end{array}
\right)\frac{(-1)^k N^{2k}r_{\rm B}^{2u-2k-1}}{2u-2k-1}\right),
\eear
\bear\label{s1}
S_{\rm Bolt, dS}&=&\frac{(4\pi)^u\beta}{16\pi l^2}
\left[-\frac{(2u-1)(2u+1)(-1)^uN^{2u+2}}{r_{\rm B}}\right.
\nonumber\\
&&\hspace{10mm}+\sum_{k=0}^{u}\left(
\begin{array}{l}
u\\
k
\end{array}
\right)(-1)^k N^{2k}r_{\rm B}^{2u-2k}
\nonumber\\&&\hspace{17mm}\times\left(-\frac{(2u-1)l^2}{(2u-2k-1)r_{\rm B}}\right.
\nonumber\\&&+\left.\left.\frac{(2u+1)(2u^2+3u-2k+1)r_{\rm B}}{(2u-2k+1)(u-k+1)}
\right)\right],\nonumber\\
\eear
where $r_{\rm B}=\frac{ql^4+\sqrt{q^2l^2+(2u+1)(2u+2)^2N^2[(2u+1)N^2+l^2]}}
{(2u+1)(2u+2)N}$.
The CFT entropy is given as
\bear
\frac{2\pi l\sqrt{|E_{\rm c}|(2E-E_{\rm c})}}{2u\sqrt{2u-1}},
\eear
where $1/\sqrt{ab}$ is fixed to $1/2u\sqrt{2u-1}$.
As the NUT charge goes to 0, the CFT entropy becomes
\bear
S_{\rm CFT}=\frac{(4\pi)^u}{4N^{2u}}
\left(\frac{ql^2}{2u^2+3u+1}\right)^{2u}=-S_{\rm Bolt,dS},
\eear
which shows that no entropy of the Taub-Bolt-dS metric satisfies the Cardy-Verlinde formula.
This means that any Bolt solution in dS space is thermodynamically unstable
at high temperature limit.

\section{Conclusion}
We have considered that the Taub-NUT/Bolt-(A)dS metric in general even dimension, and
have checked that its metric suffices to be the the Cardy-Verlinde formula.
In the limit of high temperature, we showed that the Taub-Bolt-AdS space well follows
the generalized  Cardy-Verlinde formula (\ref{cv0}) rather than
the Cardy-Verlinde formula (\ref{cv}). It seems that the modification of
the standard Cardy-Verlinde formula (\ref{cv}) is due to the distinctive property of the
Taub-NUT solution such as the ALAdS metric.
It was proven that the leading term of the CFT entropy
at the boundary for all even $u$ is exactly matched with that of the entropy
in the Taub-NUT-(A)dS space by using the generalized  Cardy-Verlinde formula
at  high temperature.
Thermal stability of the Taub-NUT-(A)dS solution for all odd $u$ and
Taub-Bolt-dS solution for all $u$ is determined by the magnitude of the NUT charge
so that the negative entropy occurs as the NUT charge goes to 0.
Finally, the the breaking of Cardy-Verlinde formula in the Taub-Bolt-dS metric
seems to reflect the fact that there is no
Bolt solution in dS space due to the absence of hyperbolic
NUT in AdS space~\cite{Mann:2004mi}.

\section*{Acknowledgements}
We are grateful to Cristian Stelea for useful comments.
This work was supported by the BK 21 project of the Ministry of
Education and Human Resources Development, Korea (C.O.L.).

\end{document}